\documentclass{icrc29}
\usepackage{graphicx,amssymb,amsmath,times}
\begin{document}

\title[Optical Relative Calibration and Stability ...]{Optical Relative Calibration
and Stability Monitoring for the Auger Fluorescence Detector}
\author[C. Aramo et al.]{C. Aramo, J. Brack, R. Caruso, D. D'Urso,  D. Fazio,
 R. Fonte, H. Gemmeke,
 \newauthor
 M. Kleifges, R. Knapik, A. Insolia, J. A. J.
Matthews, A. Menshikov, W. Miller,
\newauthor
P. Privitera and J. Rodriguez  Martino  for the Pierre Auger Collaboration$^a$ \\
(a)  Pierre Auger Observatory, Av. San Martin Norte 304, (5613)
Malargue , Argentina
 }
\presenter{Presenter: A. Insolia (antonio.insolia@ct.infn.it), \
ita-insolia-A-abs1-he15-poster }

\maketitle

\begin{abstract}
The stability of the {fluorescence telescopes} of the Pierre Auger
Observatory is monitored with the optical relative calibration
setup. Optical fibers {distribute light pulses} to three different
diffuser groups {within the optical system.} The total charge per
{pulse} is measured {for each pixel and compared with reference
calibration measurements}. This allows  monitoring the short and
long term stability with respect of the relative timing between
pixels and the relative gain for each pixel. The {designs of the LED
calibration unit (LCU)  and of the Xenon flash lamp used for
relative calibration}, are described and their capabilities to
monitor the stability of the telescope performances are studied. We
report the analysis of relative calibration data {recorded during
2004}. Fluctuations in the relative calibration constants provide a
measure of the stability of the FD.
\end{abstract}

\section{Introduction}

Data from the Pierre Auger Observatory are used to study high energy
cosmic rays in the region of the Greisen-Zatsepin-Kuz'min (GZK)
cut-off \cite{cris2004}. {The detection of events with primary
energy beyond the GZK cut-off, but also the discrepancy between
experimental data from the AGASA and HiRes experiments, results in a
wide spectrum of hypotheses about the origin of ultra high energy
cosmic rays and their nature. To solve this puzzle a well understood
energy scale and well understood energy resolution combined with
high statistics are of decisive importance for the Pierre Auger
Observatory} \cite{nim-auger}. A subset of showers observed by the
observatory is measured by both the ground array and Fluorescence
Detector (FD).
 These special {\it hybrid} events \cite{nim-auger} are
used to set the shower energy scale, based on the determination of
the shower energy by the fluorescence detector , and to measure the
shower energy resolution of the experiment. This paper will focus on
the optical relative calibration of the fluorescence telescopes and
{the accuracy of this method. Details of the absolute calibration of
FD are discussed in reference} \cite{Jeff}.

\section{Fluorescence Detector Relative Optical Calibration}
\begin{figure}[h]
\begin{center}

\includegraphics*[height=6.5cm,clip]{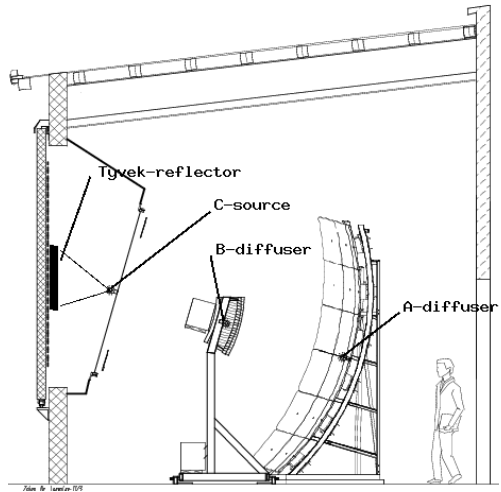}

\caption { \label {fig1} Schematic view of the Auger fluorescence
telescopes. Also shown are the positions of the diffusers for the
relative optical calibrations: A at the center of the mirror, B on
both sides of the camera and, C in the aperture box.}
\end{center}
\end{figure}

The fluorescence detectors of the Pierre Auger Observatory are
located   on top of hills at the perimeter of the ground array.
Three installations -- named Los Leones (LL), Los Morados and,
Coihueco -- are already equipped with 6 operating telescopes each
and are taking data. Fluorescence light from extensive air showers
enters the aperture of the telescope through an UV transparent
filter ($300 < \lambda < 420$ nm) \cite{matthews} and is focussed by
a spherical mirror on a camera made of 440 photomultiplier tubes
(PMTs) as shown in Fig. \ref{fig1}. Gaps between adjacent PMTs on
the focal surface are covered by reflective triangular inserts,
termed Mercedes stars. These act like Winston cones and reflect
light toward the PMTs which would otherwise be lost in the gaps.
This maximizes the light collection and results in a smooth
transition between adjacent PMTs. The diameter of the telescope was
increased from original 1.7 m to 2.2 m in the final design to
increase the effective aperture area by about a factor 2. A
corrector ring of annular shape covers the additional outer aperture
with radii of $0.85 $ m $< r < 1.1 $ m to preserve a small spot size
also at the increased diameter.

The signal of each pixel passes through a programmable gain
amplifier, an anti-aliasing filter and is digitized by a 10 MHz,
12-bit ADC  in the front-end electronics associated to each
telescope \cite{kleifges}.

The absolute calibration \cite{Jeff} provides the conversion between
the digitized signal (in ADC units) and the photon flux incident on
the 3.80 m$^2$ telescope aperture. The wavelength dependent
transparency of the UV filter, the mirror reflectivity and, the PMT
quantum efficiency results in wavelength dependent absolute
calibration constants. The electronic gain amplifier is adjusted
during the absolute calibration procedure such that the response of
the telescope to light is identical at a fixed wavelength. Changes
in the properties of any optical component would change the absolute
calibration and must be tracked in time by routine relative
calibration measurements. These measurements determine the response
of the telescope to light pulses from  either a LED source or a
xenon flash lamp. The results is compared to reference measurements
made directly after absolute calibration.  Therefore, the main goal
of the relative calibration is to monitor short term and long term
changes between successive absolute calibration measurements and to
check the overall stability of the Fluorescence Detector.

\section{The LED  and the Xenon light sources}

\underline {The LED light source}: optical fibres distribute light
from a super-bright LED array (470 nm) to the diffusers (denoted
with "A" in Fig. 1) which illuminate the cameras in the same FD
building simultaneously. Part of the LED light is directed by a
fibre to a monitoring photodiode. A LED Calibration Unit (LCU)
drives the LED array with programmable current pulses, and records
the photodiode signal.  The LCU consists of the following main
functional units: a central control, a programmable pulse generator
and a pulse monitor. The central control unit provides a TCP/IP
connection to the local LAN and controls the  operation of all LCU
functions. The programmable pulse generator is able to generate
current pulses of arbitrary shape encoded as consecutive 256 samples
of a 40 MHz 10-bit Digital-to-Analog Converter or rectangular pulses
up to $100 ~\mu$s long with programmable amplitude ($<1$ A). The
pulse monitor consists of a photodiode (Hamamatsu S1336 with zero
sensitivity drift), a transimpedance amplifier, an anti-aliasing
filter of 3.1 MHz cut-off frequency and a 10 MHz 12-bit ADC. The rms
noise (sigma) of the monitor circuit is 2.5 ADC counts. The LCU is
controlled during relative calibration by the DAQ. It defines (via
TCP-IP) the total number of light pulses, the pulse shape, its
amplitude and its duration. The LCU can be triggered by its internal
timer or by an external trigger signal. The light pulse is generated
synchronous to the 10 MHz sampling clock of the front-end
electronics with low jitter. The pulse monitor stores 1000
consecutive ADC samples from the monitoring diode for later readout
by the DAQ process.

\underline{The Xenon light source}: two xenon flash lamps are
coupled to optical fibers to distribute light pulses to different
destinations (denoted "B" and "C" in Fig. 1) on each telescope.
Source ``B'' pulses terminate at 1 mm thick Teflon diffusers on both
sides of the camera and are directed towards the mirror. Source "C"
pulses are fed through ports on the sides of the aperture and are
directed onto a reflective TYVEK foil mounted on the inner side of
the telescope doors. The foil reflects the light back into the optic
of the telescope. In this paper we focus on the results of relative
calibration with the LED source. For details on the xenon flash lamp
calibration as well as for technical references see \cite{matthews}.
We simply mention that monitoring the Los Leones and Coihueco xenon
light source stability results in mean light intensity distribution
characterized by a relative uncertainty smaller than $0.5\%$. The
two relative calibration light sources are therefore quite stable
and provide  a good tool for monitoring the long term stability of
the Fluorescence Detector.

\section{Monitoring the Fluorescence Detector with the Relative Optical Calibration}

\begin{figure}[h]
   \begin{center}

   \includegraphics*[height=10cm,width=6cm,angle=-90,clip]{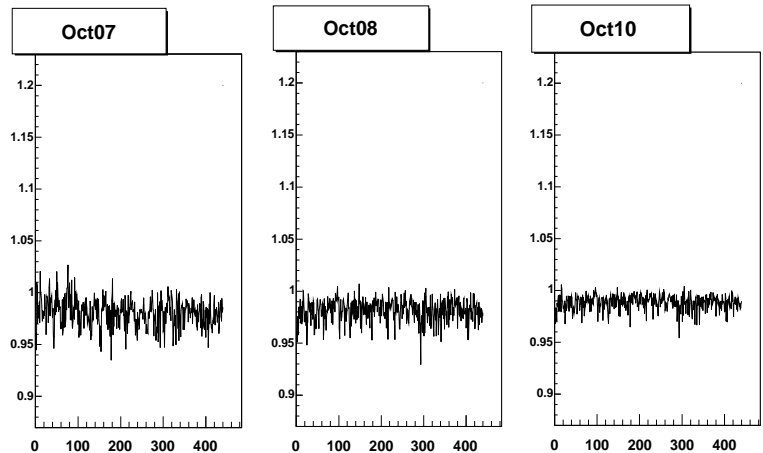}

   \caption{ \label{fig2} Typical relative calibration constants from
calibration A versus the telescope pixel number 1 to 440. The
constants fluctuate around 1.0 for all days of the October shift
(October, 7-26).}

   \end{center}
   \end{figure}

A-, B- and C- calibrations are routinely taken twice per data taking
shift. As the light sources were essentially constant, the A-source
calibration signals in each pixel (of each telescope) provided an
optimal monitor of the pixel stability.  The relative calibration
constants are measured with respect to a given reference relative
calibration run taken within one hour after the absolute calibration
measurement.  The total integrated charge is then measured with
respect to this run. A typical result for telescope \# 4 in Los
Leones is shown in Fig. \ref{fig2}.

\begin{figure}[!ht]
\begin{minipage}{7.5cm}
\begin{flushleft}
\includegraphics*[height=7cm,width=7cm,clip]{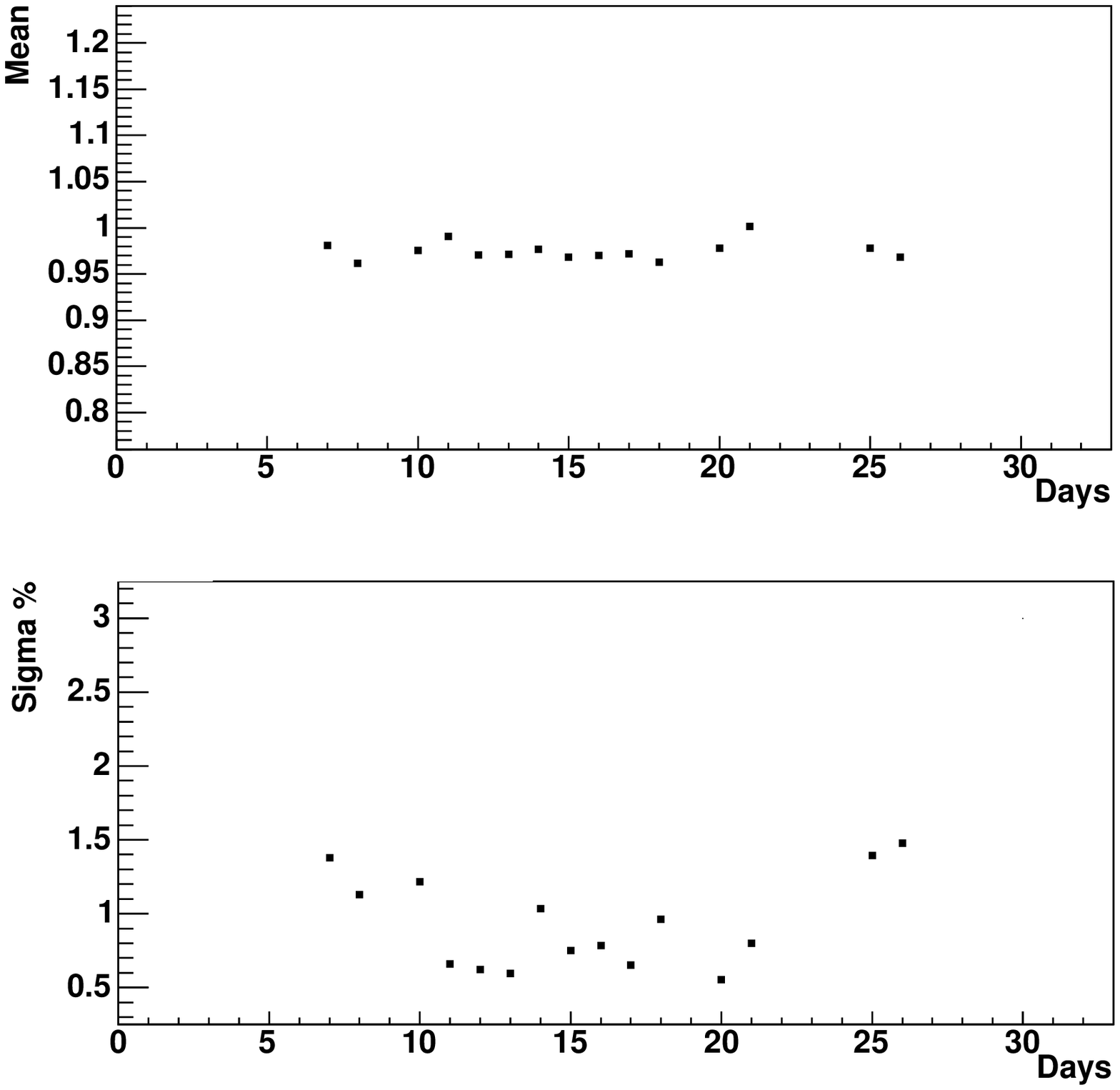}
 \caption{ \label{fig3} Daily averaged relative calibration
constants from calibration A for telescope \#2 in Los Leones during
the measuring period in October 2004.}

\end{flushleft}
\end{minipage}
\hspace{0.1cm}
\begin{minipage}{7.5cm}
\begin{flushright}
\includegraphics*[height=7.3cm,width=7cm]{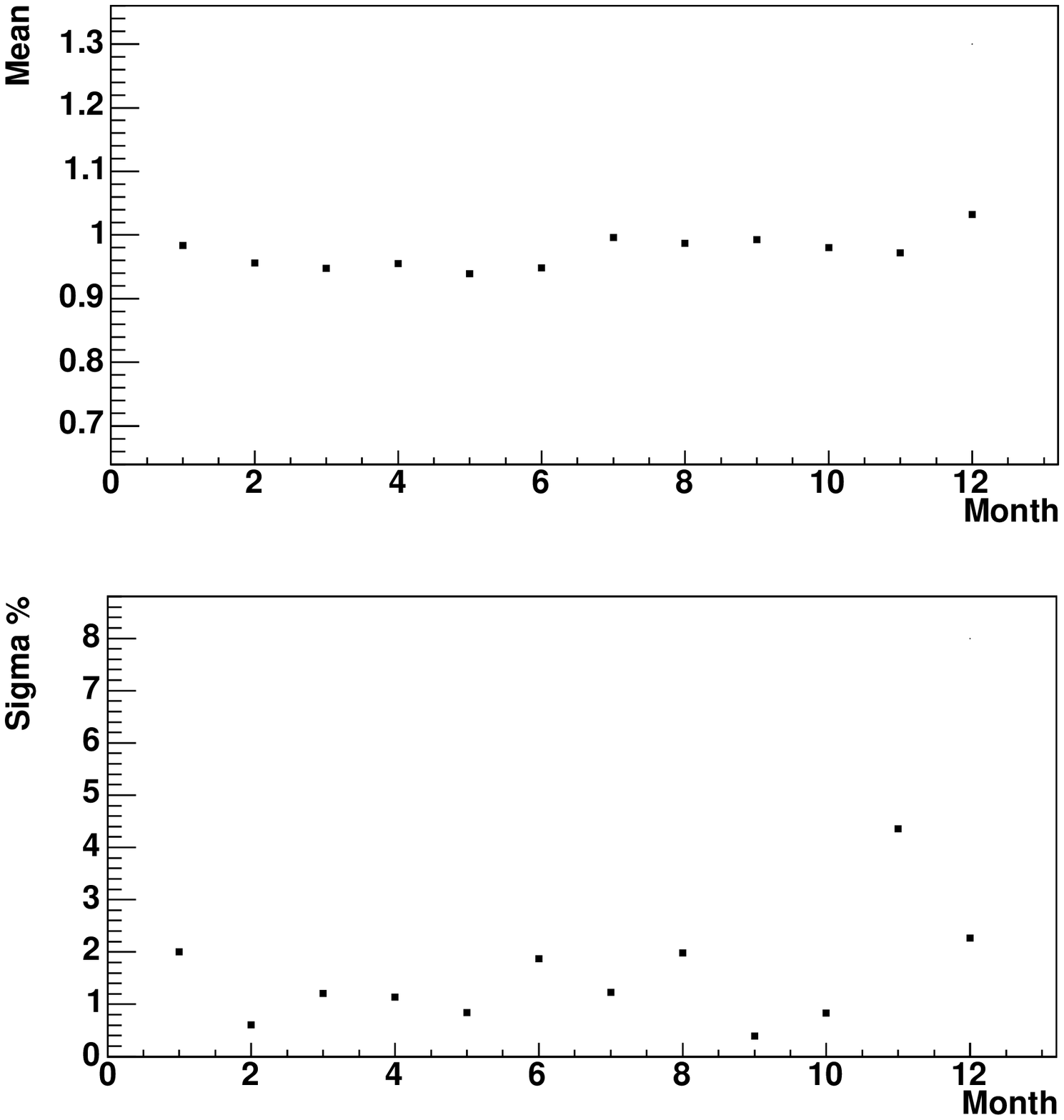}

   \caption
   { \label{fig4}
Monthly averaged relative calibration constants from calibration A
for telescope \#4 in Los Leones during the year 2004.}

\end{flushright}
\end{minipage}
\end{figure}

The relative calibration constants fluctuate around 1 with a typical
width of a few percent. To monitor the short term stability of the
system we calculated the camera averaged relative calibration
constants and its rms every day. The trend over the measuring period
in October 2004 is shown in Fig.  \ref{fig3} for telescope \# 2 of
Los Leones as an example.

The rms reported in Fig. \ref{fig3} (bottom diagram) is typical for
all relative calibration runs, resulting in a rms of about $2\%$.
The stability over a year or even longer times can be monitored by
calculating monthly averages and displaying the trend for each
telescope over the full data taking period. The data of telescope
 \# 4 in Los Leones shown in Fig. \ref{fig4} represent a typical
situation. The system is stable within a few percent both on the
long term and on a monthly base. The overall uncertainty, as deduced
from the long term monitoring of the system is, typically, in the
range of 1 to 3 \%. The relative calibration is also a very
effective tool to find problems of any kinds with the PMTs' response
or in the electronics. In addition, effects on the telescopes
sensitivity caused by work on the hardware or software show up
immediately in daily relative calibrations. Thus, any variation of
the entire FD response due to the intervention on the system can be
monitored and cross-checked.

\end{document}